\RequirePackage{fix-cm}
\documentclass[smallextended]{svjour3}       % onecolumn (second format)
\smartqed  % flush right qed marks, e.g. at end of proof
\usepackage{graphicx}
\usepackage{natbib}
\setcitestyle{aysep={}}

\usepackage{amsmath}
\usepackage{bm,booktabs,color,xcolor,siunitx}
\usepackage{amssymb}
\usepackage{amsfonts}
\usepackage{multirow}
\usepackage{rotating}
\usepackage{slashbox}
\RequirePackage[OT1]{fontenc}
\RequirePackage[colorlinks,citecolor=blue,urlcolor=blue]{hyperref}
\usepackage{algorithm}
\usepackage{algorithmic}
\usepackage{boxedminipage}
\usepackage{epstopdf}
\usepackage{threeparttable}
\usepackage{adjustbox}
\usepackage{doi}
\usepackage{url}
\usepackage{hyperref}
\usepackage{caption}
\usepackage{subcaption}
\usepackage{graphicx,psfrag,epsf}
\usepackage{enumerate}
\newcommand{\xib}{\mbox{\boldmath{$\bf \xi$}}}
\newcommand{\NN}{\mbox{\boldmath{$\bf N$}}}
\newcommand{\CC}{\mbox{\boldmath{$\bf C$}}}
\newcommand{\1}{\mbox{\boldmath{$\bf 1$}}}
\newcommand{\0}{\mbox{\boldmath{$\bf  0$}}}
\newcommand{\M}{\mbox{\boldmath{$\bf M$}}}
\journalname{Statistical Papers}
\begin{document}

\title{Phase I Distribution-Free Control Charts for Individual Observations Using Runs and Patterns
%\thanks{}
}
% Grants or other notes about the article that should go on the front
% page should be placed within the \thanks{} command in the title
% (and the %-sign in front of \thanks{} should be deleted)
%
% General acknowledgments should be placed at the end of the article.



\author{Tung-Lung Wu         %\and
	%Second Author %etc.
}


\institute{Tung-Lung Wu \at
	Department of Mathematics and Statistics, Mississippi State University, MS 39759, United States \\
	Tel.: 662-325-3414\\
	Fax:  662-325-0005\\
	\email{tw1475@msstate.edu}           %  \\
	%\and
	%S. Author \at
	%   second address
}

\date{Received: date / Accepted: date}

\maketitle

\begin{abstract}
Phase I distribution-free runs- and patterns-type control charts are proposed for monitoring the unknown target value (or location parameter) for both continuous and discrete individual observations. Our approach maintains the nominal in-control signal probability at a prescribed level by employing the finite Markov chain imbedding technique combined with random permutation and conditioning arguments. To elucidate the methodology, we examine two popular runs- and patterns-type statistics: the number of success runs and  the scan statistic. Numerical results indicate that the performance of our proposed control charts is comparable to that of existing Phase I nonparametric control charts for individual observations.
\keywords{Distribution-free \and control charts \and runs and patterns \and number of success runs
	\and scan statistic \and finite Markov chain imbedding}
 \subclass{62P30}% \and MSC code2 \and more}
\end{abstract}

\section{Introduction}
 Statistical process control (SPC) is an effective technique for monitoring the characteristics of a process. In a typical control chart application, the analysis is divided into two phases: Phase I and Phase II. Phase I control charts are employed to verify that the process is in control (IC), based on a retrospective analysis of historical data to assess process stability. Once stability is confirmed, the parameter estimates derived from Phase I data are used to establish a Phase II control chart. The primary objective of the Phase II control chart is to detect any process changes, such as shifts in location or scale, as quickly as possible.

 In a typical control chart, the plotted data points are assumed to follow a parametric distribution $F_0$, such as the normal distribution. A challenge arises when the data are not normally distributed. If the distributional assumption is not satisfied, the promised characteristics of the control chart, such as the IC average run length (ARL) or the false alarm probability (FAP), can no longer be considered reliable. In other words, these control charts are not robust to small deviations from the assumed underlying distribution. Consequently, numerous distribution-free or nonparametric control charts have been developed over the past two decades; a comprehensive review of nonparametric control charts is provided by \cite{Chakraborti-2001}.
 
 In the past decade, nonparametric or distribution-free control charts have received considerable attention in quality and process control. The mainstream approaches include rank-based control charts \citep{Zhou-2009,Hawkins-2010} and likelihood-based control charts \citep{Ning-2015}. Typically, there are three types of control charts: exponentially weighted moving average (EWMA), cumulative sum (CUSUM), and Shewhart control charts. Conventional Shewhart control charts can be enhanced by incorporating supplemental runs rules. EWMA and CUSUM charts are particularly popular because they are more effective at detecting small mean shifts. For example, nonparametric EWMA control charts have been developed in the works of \cite{Amin-1991} and \cite{Zou-2010}, while nonparametric CUSUM control charts have been investigated by \cite{Chatterjee-2009}, \cite{Chowdhury-2015}, and \cite{Li2013}.
 
 The simplicity, ease of implementation, and tractability of run-length characteristics make Shewhart-type charts highly attractive. Runs rules have been applied to enhance Shewhart control charts for detecting small mean shifts, similar to their application in scan statistics \citep{Shmueli-2003}. However, the performance of these charts is not guaranteed when normality or specific parametric models are assumed \citep{Charles-1987,koutras-2007}. To overcome these limitations, \cite{Chakraborti-2009} proposed a Phase-II nonparametric control chart based on precedence statistics with runs rules for monitoring an unknown location parameter. Their method signals when two consecutive plotting points—such as the median—fall outside the control limits. Another approach, based on the Wilcoxon signed-rank statistic combined with runs-type rules, was presented by \cite{Chakraborti-2007}; however, this statistic is applicable only when the median is known. Finally, \cite{Balakrishnan-2010} introduced a general class of nonparametric control charts based on order statistics, providing a detailed review of the field.

 The Phase-I control chart is crucial in process monitoring, as the success of the Phase-II control chart depends on the effectiveness of the Phase-I control chart. However, there are limited results in the literature regarding Phase-I control charts. Over the last two decades, most distribution-free or nonparametric control charts have been designed for Phase II. Consequently, there is a need for new Phase-I control charts. In addition, nonparametric Shewhart control charts with runs rules appear to be underdeveloped relative to nonparametric EWMA and CUSUM control charts. One likely reason is the lack of methods for handling general runs without any distributional assumptions. To bridge these gaps, we develop Phase-I distribution-free control charts based on runs and patterns.
 
%Recent works have also addressed the challenges specific to Phase I control charting, including the treatment of retrospective data, subgroup formation, and parameter estimation uncertainty (Jones-Farmer et al. 2009, 2014; Capizzi 2015). These studies emphasize the need for robust, flexible tools capable of handling complex Phase I data structures. The proposed runs- and scan-based distribution-free control charts complement this literature by providing exact, model-free charting procedures for individual observations. Our approach focuses on the fundamental probabilistic properties of runs and scan statistics, which can be further incorporated into broader Phase I frameworks as discussed in the aforementioned studies.

 Recent research has emphasized the distinct challenges of Phase I analysis, including the treatment of retrospective data, estimation uncertainty, subgroup formation, and distributional assumptions. In particular, \cite{Jones-Farmer-2009,Jones-Farmer-2014} and \cite{Capizzi-2015} have provided valuable frameworks for addressing such issues and for integrating estimation and diagnostic considerations into Phase I charting. Our proposed runs- and scan-based control charts complement this literature by providing exact distribution-free methods for individual observations. In contrast to existing Phase I approaches that rely on estimated parameters or subgroup summaries, the present work establishes a model-free foundation using conditional distributions of runs and scan statistics derived via the finite Markov chain imbedding technique.

 The finite Markov chain imbedding (FMCI) technique has been employed to evaluate the exact characteristics of the run-length distribution, such as the mean and standard deviation, of various Phase II control charts, including Shewhart, EWMA, and CUSUM charts under the assumption of normality \citep{Fu-2002,Fu-2003}. In this work, we study Phase I distribution-free runs- and patterns-type control charts, with the conditional distributions of runs and patterns derived exactly using the FMCI technique. In this framework, Bernoulli trials, given the total number of successes, are treated as random permutations. This approach coincides with other works which have shown that the conditional distributions of the proposed monitoring statistics are independent of unknown parameters and the underlying distributions \citep{Chen-2016}.
 
 %Three classic control charts are Shewhart,EWMA  and  CUSUM control charts.  It is well-known that Shewhart control charts can quickly detect large shits while the EWMA and CUSUM control chart are better for small shifts.
 %In practice, we prefer the Shewhart control chart since it is easy to understand and interpret by practitioners. Many supplemented rules are developed to improve the ability of the Shewhart control charts in detecting small shifts. Some well-known runs statistics are the longest run and scan statistic. In this manuscript, we will construct distribution-free charting procedures for these two runs statistics and study their properties. 
 
 Much of the existing literature on distribution-free runs-type control charts is either limited to specific run statistics or not entirely distribution-free. In this study, we propose a general framework for constructing distribution-free runs- and patterns-type control charts for individual observations.

 The remainder of this manuscript is organized as follows. Section~2 introduces the conditional distributions of the number of success runs, $R_n$, and the scan statistic, $S_n(r)$. Section~3 outlines the charting procedures for  scan-based and runs-based control charts. Section~4 presents the numerical results. Section~5 demonstrates an application using piston ring data. Finally, Section 6 offers concluding remarks and discussion.
 
 \section{Conditional distributions of $R_n$ and $S_n(r)$ given the number of successes}
 
  Let $X_1,\ldots,X_n$ be a sequence of independent and identically distributed (i.i.d.) Bernoulli trials with success probability $p=\Pr(X_1=1)$, and let $N_1$ denote the number of successes in the sequence. %We want to compute the conditional probabilities $\Pr(R_n<r|N_1=n_1)$ and $\Pr(S_n(r)<s|N_1=n_1)$. 
 This section provides a detailed account of the finite Markov chain imbedding (FMCI) technique, which is employed to derive the exact conditional distributions of the number of success runs and scan statistics given the number of successes. 
 
 \subsection{Number of success runs}
 A \emph{success run} is defined as a maximal consecutive subsequence of 1s, that is, 
 a sequence $(X_j, X_{j+1}, \ldots, X_{k}) = (1,1,\ldots,1)$ such that $X_{j-1}=0$ (or $j=1$) and $X_{k+1}=0$ (or $k=n$). 
 The total number of such success runs in the sequence is denoted by $R_n$.
 
 The exact \textit{unconditional} distribution of the number of success runs was derived by \cite{Fu-1994}. The \textit{conditional} distribution of the number of success runs has been discussed in several studies, including \cite{Lou-1997} and the book by \cite{Fu-lou-2003}. In this work, we adopt the method presented by \cite{Lou-1997}, as it leads to a smaller state space for the imbedded Markov chain. Let  
 \[
 \mathcal{P} = \left\{\boldsymbol{\pi} = (\pi_1, \ldots, \pi_n) : \pi_i \in \{0,1\}, \sum_{i=1}^n \pi_i = n_1 \right\}
 \]  
 denote the set of random permutations consisting of \( n_1 \) ones and \( n_2 = n - n_1 \) zeros. The conditional distribution of \( R_n \), given a total of \( n_1 \) successes in a sequence of \( n \) Bernoulli trials, is equivalent to the distribution of \( R_n \) in a random permutation \( \boldsymbol{\pi} \in \mathcal{P} \).
 
 To construct the imbedded Markov chain, we sequentially and randomly insert \( n_2 \) zeros into an initial row of \( n_1 \) ones, one at a time, over \( n_2 \) steps. At each insertion step, we record the number of success runs. The Markov chain \( \{Y_t\} \) is thus defined on the state space \( \Omega_1 = \{1, 2, \ldots, d\} \), where \( d = \min(n_1, n_2 + 1) \) is the maximum possible number of success runs. During the insertion process, a new success run is created if a 0 is inserted between two consecutive 1s; otherwise, the number of runs remains unchanged.
 
 Let \( \M_t^1 \) denote the transition matrix at time \( t \), for \( t = 1, 2, \ldots, n_2 \). The transition probabilities are defined as  
 \begin{align}
 p_{ij}(t) &= \Pr(Y_t = j \mid Y_{t-1} = i) \nonumber\\
 &= 
 \begin{cases}
 	\frac{n_1 - i}{n_1 + t}, & \text{if } j = i + 1,\ 1 \leq i \leq t, \\
 	\frac{t + i}{n_1 + t}, & \text{if } j = i,\ 1 \leq i \leq t, \\
 	1, & \text{if } j = i \geq t + 1,\ i \leq d - 1,\ \text{or } j = i = d, \\
 	0, & \text{otherwise}.
 \end{cases}
 \end{align}
 
 According to Theorem~2.1 in \cite{Fu-lou-2003}, the conditional random variable \( R_n \) is finite Markov chain imbeddable. The conditional distribution of \( R_n \), given \( N_1 = n_1 \), is then expressed as  
 \begin{equation}\label{thm11}
 	\Pr(R_n = r \mid N_1 = n_1) = \boldsymbol{\xi}_0 \prod_{t=1}^{n_2} \M_t^1 \boldsymbol{e}_r,
 \end{equation}
 where \( \boldsymbol{\xi}_0 = (1, 0, \ldots, 0) \) is the initial distribution vector and \( \boldsymbol{e}_r \) is a column vector with a 1 in the \( r \)th position and zeros elsewhere.
 
 As a result, one can determine the lower \( \alpha \)-percentile of the distribution \( \Pr(R_n = r \mid N_1 = n_1) \), denoted by \( R_n(\alpha) \).
 
 \begin{remark}
 	Note that in an $[n - n_1, n_1]$-specified random permutation, the distribution of $R_n$ is independent of the success probability $p$.
 \end{remark}
 
 An example is given below to illustrate the method. 
 
 \begin{example} 
 	Let $n=5$ and $n_1=3$. Then, $n_2=2$ and $d = \min(n_1, n_2+1) = 3$. Suppose a realization of the Bernoulli sequence is 10101; in this case, the number of success runs is $R_5=3$. For another realization, say 01110, we have $R_5=1$. 
 	
 	To compute, for example, $\Pr(R_5=2 \mid N_1=3)$, we define the state space $\Omega_1 = \{1, 2, 3\}$, the initial distribution $\boldsymbol{\xi}_0 = (1, 0, 0)$, and the unit vector $\mathbf{e}_2 = (0, 1, 0)^\top$. The transition matrices are given by:
 	\[
 	\M^1_1 = \begin{bmatrix}
 		0.5 & 0.5 & 0 \\
 		0 & 1 & 0 \\
 		0 & 0 & 1 
 	\end{bmatrix}, \quad
 	\M^1_2 = \begin{bmatrix}
 		0.6 & 0.4 & 0 \\
 		0 & 0.8 & 0.2 \\
 		0 & 0 & 1 
 	\end{bmatrix}.
 	\]
 	Then, it follows from Equation~(\ref{thm11}) that
 	\[
 	\Pr(R_5=2 \mid N_1=3) = \boldsymbol{\xi}_0 \left( \prod_{t=1}^{2} \M^1_t \right) \mathbf{e}_2 = 0.6.
 	\]	
 \end{example} 
 
 \subsection{Scan Statistics}
 
 The scan statistic is defined as \begin{equation} S_n(r) = \max_{1 \leq t \leq n - r + 1} S_n(r, t), \end{equation} where $S_n(r, t) = \sum_{i = t}^{t + r - 1} X_i$, and $r$ is the window size.
 \cite{Fu-Lou-Wu-2012} employed the finite Markov chain imbedding (FMCI) technique to derive the exact conditional distributions of scan statistics.
 
 Consider a sequence of Bernoulli trials with binary outcomes $\{0, 1\}$. Let $\Lambda = \bigcup_{i=1}^{L} \Lambda_i$ represent a compound pattern composed of $L$ simple patterns $\Lambda_1, \Lambda_2, \ldots, \Lambda_L$, where each simple pattern $\Lambda_i$ is a fixed sequence over the binary alphabet $\{0, 1\}$ with a specified length. When any of the simple patterns occurs, the compound pattern is said to occur.
 Let $W(\Lambda)$ denote the waiting time until the first occurrence of the compound pattern $\Lambda$ in the sequence. The distribution of the scan statistic is then determined using the FMCI technique by using the dual relationship between the distribution of the scan statistic and the waiting time distribution of the associated compound pattern.
 An example is given below.
 
 \begin{example} 
 Let $r = 5$ and $s = 2$. Thus, the event $\{S_n(5) < 2\}$ means that in every window of five consecutive trials, there is at most one success.  
 This correspondence reflects a duality between the scan statistic and local pattern occurrences.  
 Specifically, the event $\{S_n(5)<2\}$ means that every window of length~5 contains at most one success, which is equivalent to requiring that no two 1s appear within fewer than~5 positions of each other.  
 Hence, the sequence must avoid all the patterns $\{11\}$, $\{101\}$, $\{1001\}$, and $\{10001\}$.  
 These four patterns form the compound pattern $\Lambda_{5,2}$, so that the probability $\Pr(S_n(5)<2)$ coincides with the probability that none of the patterns in $\Lambda_{5,2}$ occurs within the first~$n$ trials, i.e.,
 \[
 \Pr(S_n(5)<2) = \Pr(W(\Lambda_{5,2})>n),
 \]
 where $W(\Lambda_{5,2})$ denotes the waiting time until the first occurrence of any pattern in $\Lambda_{5,2}$.  
 In the conditional case, the dual relationship between the scan statistic and the compound-pattern waiting time still holds. Specifically,
 \[
 \Pr\left(S_n(5) < 2 \,\middle|\, N_1 = n_1\right) = \Pr\left(W(\Lambda_{5,2}) > n \,\middle|\, N_1 = n_1\right).
 \]
 \end{example} 
 
 Given $r$ and $s$, a set of simple patterns of lengths no longer than $r$ is defined as
 \[
 \Lambda_{r,s} = \{\Lambda_i : i = 1, \ldots, \ell \},
 \]
 where each $\Lambda_i$ is a simple pattern that begins and ends with $1$ and contains a total of $s$ ones, and $\ell$ is the total number of simple patterns corresponding to the scan statistic with parameters $r$ and $s$. 
 For each scan statistic probability $\Pr(S_n(r)<s)$, the total number of simple patterns is
 \[
 \ell = 
 \sum_{\nu = 0}^{r - s} \binom{s - 2 + \nu}{\nu}.
 \]
 Let
 \[
 \mathcal{P} = \left\{ \boldsymbol{\pi} = (\pi_1, \ldots, \pi_n) : \pi_i \in \{0,1\} \text{ and } \sum_{i=1}^{n} \pi_i = n_1 \right\}
 \]
 be the family of random permutations with $n_1$ ones and $n - n_1$ zeros. 
 Then, the conditional distributions of runs and patterns, given the total number $n_1$ of successes in a sequence of $n$ Bernoulli trials, are the same as the distributions of runs and patterns in an $[n - n_1, n_1]$-specified random permutation $\boldsymbol{\pi} = (\pi_1, \ldots, \pi_n)$. 
 It is notable that, in an $[n - n_1, n_1]$-specified random permutation, the distributions of runs and patterns are independent of $p$.

 To construct the imbedded Markov chain, a similar insertion process is used and explained as follows. To represent the sequential sampling process, imagine an urn containing $n_1$ balls labeled 1 and $n_0=n-n_1$ balls labeled 0. Balls are drawn one by one without replacement until $\Lambda_{r,s}$ occurs or the urn is emptied. Each draw updates the partial sequence, and we record the current number of observed 1s together with the longest subpattern that has appeared so far. Let $E_{r,s}$ denote the collection of all subpatterns of $\Lambda_{r,s}$ (excluding the patterns themselves), known as \emph{ending blocks}, and let $\mathcal{S}_2 = \{0,1\}$.
  Then, the nonhomogeneous Markov chain $\{Y_t\}_{t=0}^n$ can be defined on the state space   
 $$\Omega_2 = \{(m,\omega): m = 0,1,
 \ldots,n_1 \mbox{ and } \omega \in E_{r,s}\cup \mathcal{S}_2\}\cup\{\emptyset,\alpha\},$$
 where $\emptyset$ is the initial state and $\alpha$ is the absorbing state, meaning that the compound pattern $\Lambda_{r,s}$ occurs once the chain enters the absorbing state. 
 
To make the transition probabilities in Equation~(\ref{tran}) fully explicit, we first fix notation for the imbedded Markov chain $\{Y_t\}_{t=0}^n$. 
Let $\boldsymbol{\pi} = (\pi_1, \dots, \pi_n)$ denote an $[n - n_1, n_1]$–specified random permutation, where each $\pi_t \in \{0,1\}$ represents the outcome at time~$t$. 
For $t = 0, 1, \dots, n$, define
\[
m_t = \sum_{i=1}^t \pi_i, \qquad m_0 = 0,
\]
as the cumulative number of ones observed up to time~$t$. 
Recall that $E_{r,s}$ is the set of all nonabsorbing subpatterns and $\mathcal{S}_2 = \{0,1\}$; we write $\omega_t \in E_{r,s} \cup \mathcal{S}_2$ for the current subpattern after observing $\pi_t$. 
For two strings $\omega$ and $b \in \{0,1\}$, we define the concatenation operator $\langle \omega, b \rangle$ as the word obtained by appending $b$ to $\omega$, and let
\[
\langle \omega, b \rangle_{E_{r,s}}
\]
denote the longest subpattern of this concatenation that belongs to $E_{r,s}$ (or the absorbing state $\alpha$ if the concatenation completes a compound pattern). 
With these definitions, the three cases in Equation~(\ref{tran}) correspond respectively to  
(i) observing a~1 and incrementing the count $m_t = m_{t-1} + 1$ while updating the subpattern to the corresponding longest subpattern;  
(ii) observing a~0 and keeping the count unchanged while updating the subpattern; and  
(iii) staying in the absorbing state $\alpha$ once it has been entered.

 Let $\ell_{n_1}+1$ denote the size of the state space $\Omega_2$. The  transition probabilities from state $u=(m_{t-1},\omega_{t-1})$ to  state $v=(m_t,\omega_{t})$ are 
 \begin{align}
 	p_{uv}(t) &= \Pr(Y_t = (m_{t},\omega_{t})|Y_{t-1}=(m_{t-1},\omega_{t-1})),\nonumber\\
 	&= \left\{\begin{array}{lll}\label{tran}
 		\frac{n_1-m_{t-1}}{n-t+1} & \text{if} & \pi_t=1, m_t=m_{t-1}+1 \text{ and } \omega_t=<\omega_{t-1},1>_{E_{r,s}},\\[4pt]
 		\frac{n-n_1-t+m_{t-1}+1}{n-t+1} & \text{if} & \pi_t=0, m_t=m_{t-1} \text{ and } \omega_t=<\omega_{t-1},0>_{E_{r,s}},\\[4pt] 
 		1 & \text{if} & \omega_t =\omega_{t-1}= \alpha,\\
 		0 & \text{if} & \text{otherwise}.\\[4pt]
 	\end{array}\right. 
 \end{align}
 Therefore, the transition matrices are of the form
 \begin{equation}\label{Eq5}
 	\M^2_t =\left[ \begin{tabular}{c|c}
 		$ \NN^2_t$  & $ \CC^2_t$ \\ \hline
 		$ \0$  &  $1$
 	\end{tabular} \right]_{(\ell_{n_1}+1)\times(\ell_{n_1}+1)}, %\label{eq:2.1}
 \end{equation}
 where $\NN^2_t,t=1,2,\ldots,$ are $\ell_{n_1}\times\ell_{n_1}$ submatrix of transition probabilities among transient states, and 
 $\CC_t^2$ is an $\ell_{n_1}\times 1$ column vector whose $i$th entry gives the probability that the chain moves from  state~$i$ into the absorbing state~$\alpha$.
 
 Since we are interested in the event that the waiting time exceeds $n$ (i.e., the chain remains within the transient states associated with $\NN^2_t$ up to time~$n$), the conditional waiting-time probability can be obtained from Theorem~2.1 of \cite{Fu-lou-2003}. 
 Hence, the conditional distribution of $S_n(r)$ given $N_1 = n_1$ is
 \begin{align}\label{cond}
 	\Pr\!\left(S_n(r)<s \,\middle|\, N_1 = n_1\right)
 	&= \Pr\!\left(W(\Lambda_{r,s}) > n \,\middle|\, N_1 = n_1\right) \nonumber\\
 	&= \xib_0 \Bigg(\prod_{t=1}^n \NN^2_t\Bigg) \1^{\top},
 \end{align}
 where $\xib_0$ is a $1\times\ell_{n_1}$ vector of initial probabilities, $\NN^2_t,t=1,\ldots,n,$ are $\ell_{n_1}\times\ell_{n_1}$ matrices whose entries are given in Equation~(\ref{tran}), and $\1$ is a $\ell_{n_1}\times1$ row vector of ones.  
 
 The conditional distribution in Equation~(\ref{cond}) is independent of $p$. For more details on the imbedding Markov chain, refer to  \cite{Fu-Lou-Wu-2012}.
 
 \begin{remark}
 	The two statistics, the number of success runs $R_n$ and the scan statistic $S_n(r)$, are used for illustrative purposes. In fact, the methodology introduced in this section can be used to compute the conditional distributions of general runs and patterns of fixed lengths. The construction of the transition matrix  depends on the specific runs or patterns employed. Further details can be found in \cite{Wu-2020}.  
 \end{remark}

 \section{The proposed control charts}

In this section, we propose Phase-I distribution-free runs- and patterns-type control charts that are applicable to any underlying process distribution $F_0$. The methodology for one-sided control charts, as outlined throughout this paper, is designed primarily for detecting upward location shifts. However, it can also be extended to control charts with both lower and upper limits, enabling the detection of both downward and upward shifts.

Let $\{Y_1,\ldots,Y_n\}$ be a sample from an unknown distribution $F_0$. 
In the one-sided control chart, a label of ``1'' is assigned to an observation if its value exceeds a certain threshold $c$; otherwise, it is labeled as ``0''. Specifically, we define $X_n$ as follows:
\begin{align}\label{111}
	X_n= \left\{\begin{array}{lll}
		1 & \text{if } & Y_n\geq c,\\
		0 & \text{if } & Y_n< c.\\
	\end{array}\right. 
\end{align}
The number of success runs and scan statistics can then be defined for the  sequence $\{X_n\}$, consisting of the two outcomes 0 and 1.  It has been widely established that runs rules can enhance the detection of small shifts in Shewhart-type control charts. 
In addition to the two run statistics $R_n$ and $L_n$ discussed herein, one can consider general runs and patterns rules, denoted by $R(k,r,Z)$, as  proposed by \cite{Shmueli-2003}. These rules are defined as follows: if the $k$ of the last $r$-tested points fall in the region $Z$,  the control chart signals an out-of-control (OC) alert.  $R(k,r,(c,\infty))$ and $R(k,k,(c,\infty))$ are generally referred as scan rules and run rules, respectively.\\

Given a pre-specified level $\alpha$, the charting procedure based on the number of success runs $R_n$ (denoted as R-1) is outlined below.  

\begin{algorithm}[H]
	\caption{The R-1 Control Chart}
	\begin{algorithmic}[1]
		\STATE \textbf{Initialization:} $\{Y_1, \ldots, Y_n\}, \alpha, n, p_0$
		\STATE Determine the threshold $c$ using Equation~(\ref{eq-c})
		\STATE Obtain $\{X_1, \ldots, X_n\}$ using Equation~(\ref{111}) and set $n_1 = \sum_{i=1}^{n} X_i$
		\STATE Compute the lower $\alpha$-percentile $R_n(\alpha)$ using Equation~(\ref{thm11})
		\STATE Compute the observed value of $R_n$
		\STATE Signal an alarm if $R_n \le R_n(\alpha)$
	\end{algorithmic}
\end{algorithm}

Similarly, the charting procedure based on the scan statistic $S_n(r)$ (denoted as R-2) is given in Algorithm 2. 

\begin{algorithm}[!h]
	\caption{The R-2 Control Chart}
%	\begin{boxedminipage}{155mm}
		\begin{algorithmic}[1]
			\STATE \textbf{Initialization:} $\{Y_1, \ldots, Y_n\}, \alpha, n, p_0$
	     	\STATE Determine the threshold $c$ using Equation~(\ref{eq-c})
			\STATE Obtain $\{X_1, \ldots, X_n\}$ as defined in Equation~(\ref{111}) and set $n_1=\sum^{n}_{i=1} X_i$
			\STATE Compute the upper $\alpha$-percentile $S_n(r, \alpha)$ using Equation~(\ref{cond})
			\STATE Compute the observed value of $S_n(r)$
			\STATE Signal an alarm if $S_n(r) \geq S_n(r, \alpha)$
		\end{algorithmic}
%	\end{boxedminipage}
%	\vskip17.5pt
\end{algorithm}
Note that the runs- and patterns-type statistics are discrete, and therefore, controlling the IC signal probability at any given level may not always be feasible. As a result, a randomized test is used to control the IC signal probability at the desired level. 
 
 \subsection{Choice of  $c$}
 
 The binary transformation in Equation~(\ref{111}),
 \[
 X_i = I(Y_i \geq c), \qquad i=1,\ldots,n,
 \]
 converts the observed data into a $0$--$1$ sequence, where $N_1 = \sum_{i=1}^n X_i$ denotes the number of 1s and $N_0 = n - N_1$ the number of 0s. The in-control signal probability, conditional on $N_1 = n_1$, does not depend on the underlying distribution $F_0$ or the threshold $c$, since the conditional distribution of the binary sequence is uniform over all permutations of $n_1$ ones and $n_0$ zeros.
 Because the conditional distribution of the proposed statistic depends only on $N_1$, the choice of the threshold $c$ determines the baseline proportion 
\begin{equation}\label{eq-c}
 p_0 = \Pr(Y \geq c)
\end{equation}
 and therefore influences the sensitivity of the resulting chart.
 
 The threshold $c$ may generate an excessively sparse or excessively dense sequence, leading to long strings of zeros or ones and consequently reducing the resolution of the chart. To mitigate this, $c$ is chosen so that the transformed observations yield a balanced proportion of zeros and ones. In practice, this is accomplished by selecting $c$ as the empirical quantile $\widehat q_{\,1-p_0}$ from the Phase~I reference sample, where $p_0$ is restricted to a moderate range such as $(0.1,\,0.8)$. This approach prevents degeneracy in the resulting binary sequence. The recommended values of $p_0$ are discussed in Section 4.

% For {\it rare-event} detection we select $c$ so that the exceedance rate $p_0=\Pr(Y>c)$ is small but nondegenerate.  A  choice for the scan statistic is
% p_0^\star=\sup\{p:\Pr(\mathrm{Bin}(r,p)\ge %s)\le \alpha_0\},
% \quad c=\widehat F_0^{-1}(1-p_0^\star),
% \]
% with a small tail level $\alpha_0$ (e.g., %$10^{-3}$--$10^{-2}$). 
% This keeps $s$ or more exceedances in an %$r$-window rare under in-control conditions, %enhancing cluster sensitivity.  
% In Phase~I, $c$ is computed from the %empirical quantile $\widehat F_0^{-1}$ of the %reference sample; we recommend a short %sensitivity check over nearby $p_0$ values %(e.g., $0.05,0.08,0.10$). 

 \subsection{Identification of potential sources of process instability}
 When the chart signals in Phase~I, we report explicit localization outputs so that the practitioner can identify \emph{where} instability occurs.
 
 \paragraph{Scan statistic (window localization).}
 Let $X_i=I(Y_i\geq c)$ and fix the window length $r$.
 For each window $I_j=\{j,\ldots,j+r-1\}$, define the count $C_j=\sum_{i\in I_j} Y_i$.
 Upon a signal, we return the top-$K$ windows
 \[
   I_{\hat{\jmath}_1},I_{\hat{\jmath}_2},\ldots,I_{\hat{\jmath}_K} \ \text{such that}\  C_{\hat{\jmath}_1} \;\ge\; C_{\hat{\jmath}_2} \;\ge\; \cdots \;\ge\; C_{\hat{\jmath}_K},
 \]
 together with their conditional $p$-values computed under the distribution-free model conditional on $N_1$.
 The center of $I_{\hat{\jmath}_1}$ is reported as an anomaly location proxy $\hat\tau$.
 We list the time indices within each of the top-$K$ windows to support root-cause investigation.

 \paragraph{Success-run statistic (run localization).}
 Let $R_1\ge R_2\ge \cdots$ denote the ordered lengths of consecutive runs of 1s in $\{X_i\}$.
 Upon a signal, we return the indices of the longest (or top-$K$ longest) runs, their start and end positions, and conditional $p$-values.
 These segments pinpoint locations where extended success episodes arise.

  .
 
 \section{Numerical Results and Comparison}
 
 The proposed control chart is compared with existing ones based on a widely adopted change-point model \citep{Zhou-2009, Parpoula-2021}. The process monitoring in Phase-I is analogous to the change-point detection problem, specifically for detecting a location shift. When the process is IC, we assume that the individual observations $X_1,X_2,\ldots,X_n$ are sampled independently from an unknown distribution $F_0$ with unknown mean $\mu_0$ and variance $\sigma^2_0$. When the process is OC, we assume the data are drawn from the following model:  
\begin{align}\label{model}
	X_t = 
	\begin{cases}
		F_0(x), & \text{for } t = 1, \ldots, \tau, \\
		F_1(x), & \text{for } t = \tau + 1, \ldots,
	\end{cases}
\end{align}
 where $\tau$ is the unknown change point. 
 
The IC signal probability is set to 0.005, and three distribution functions are considered: the standard normal distribution, the exponential distribution with a mean of 1, and the $t$-distribution with 3 degrees of freedom. 
These distribution choices cover the cases of symmetric, skewed, and heavy-tailed distributions.
 
 To better compare the performance of the proposed control charts with existing Phase-I control charts, we follow the settings in \cite{Parpoula-2021}, where a step change in the process mean occurs when the process becomes OC. Let $\mu_j=E(X_j)$, with $\mu_1=\mu_2=\cdots=\mu_\tau=\mu_0$ and $\mu_{\tau+1}=\mu_{\tau+2}=\cdots=\mu_n=\mu_1=\mu_0+\delta\sigma_0$, for an unknown time point $\tau$. Three OC scenarios are
 considered:
 \begin{description}
 	\item[Scenario I:] $\tau=10$ for $n=50$ and $\tau=20$ for $n=100$; 
 	
 	\item[Scenario II:] $\tau=25$ for $n=50$ and $\tau=50$ for $n=100$; 
 	
 	\item[Scenario III:] $\tau=40$ for $n=50$ and $\tau=80$ for $n=100$. 
 \end{description}
 The three scenarios represent distinct patterns: (I) the OC period is longer than the IC period; (II) the OC period is equal to the IC period; and (III) the OC period is shorter than the IC period.
 
We first examine the effect of the reference proportion $p_0$ (or equivalently, $c$) on the performance of the proposed control charts. 
 In Phase~I applications, the threshold $c$ should be selected so that an excessive number of successes is primarily due to the instability.
To illustrate this, we adopt a simple \emph{quantile-targeting rule} that sets $c$ according to a desired target success rate $p_0 = 0.2$, $0.5$, or $0.8$.  Let $\hat{q}_{1-p_0}$ be the empirical $(1-p_0)$ quantile of the Phase~I reference sample; we then define
\[
c = \hat{q}_{1-p_0},
\qquad
\hat{p}_0 = \frac{1}{n}\sum_{i=1}^{n} I(Y_i \geq c).
\]

\begin{figure}[!h]
 	\centering
 	\includegraphics[scale=.44]{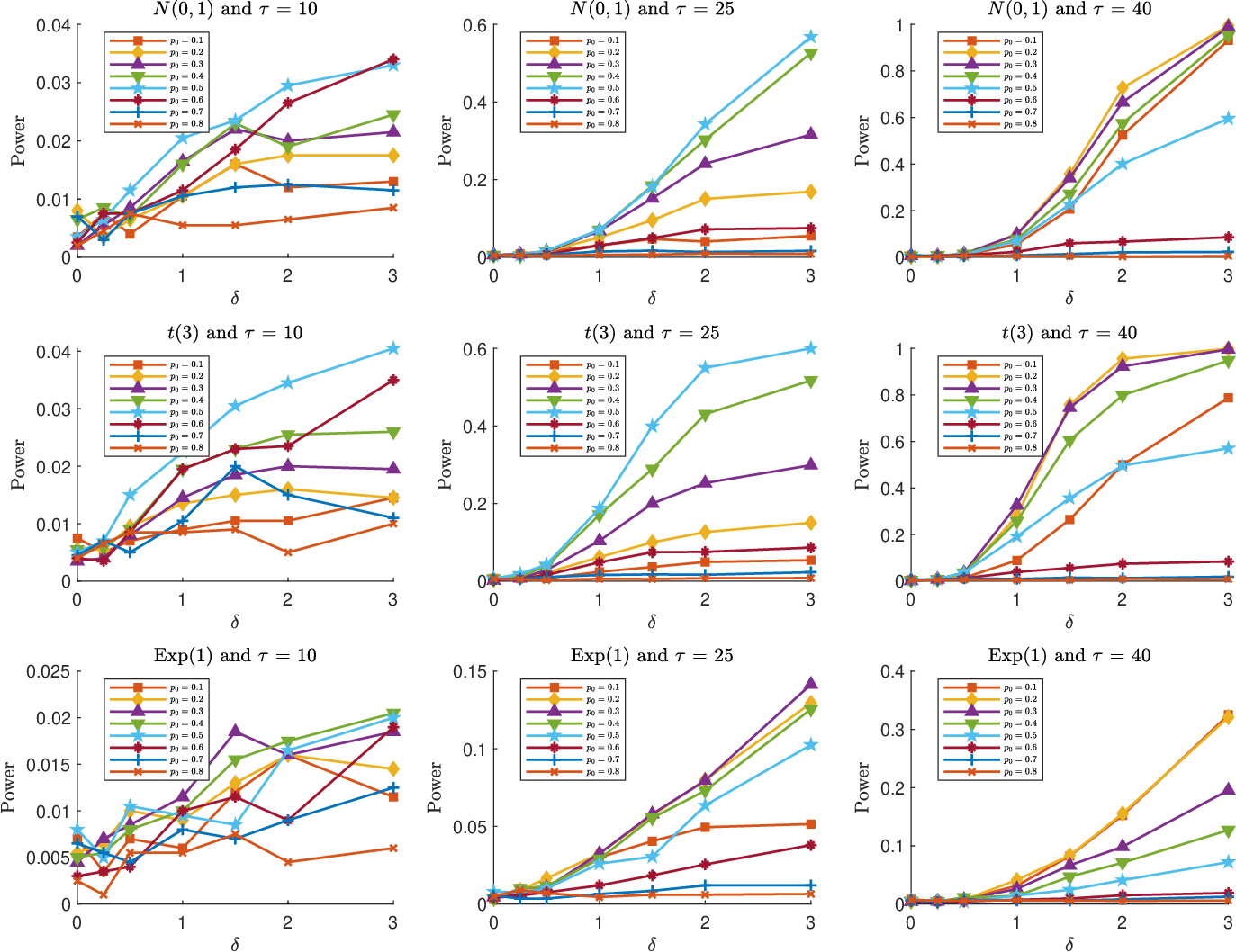}
 	\caption{OC signal probabilities for the R-2 chart with $n=50$ and window size $r=10$.}\label{f1}
 \end{figure}

\begin{figure}[!h]
 \centering
 \includegraphics[scale=.44]{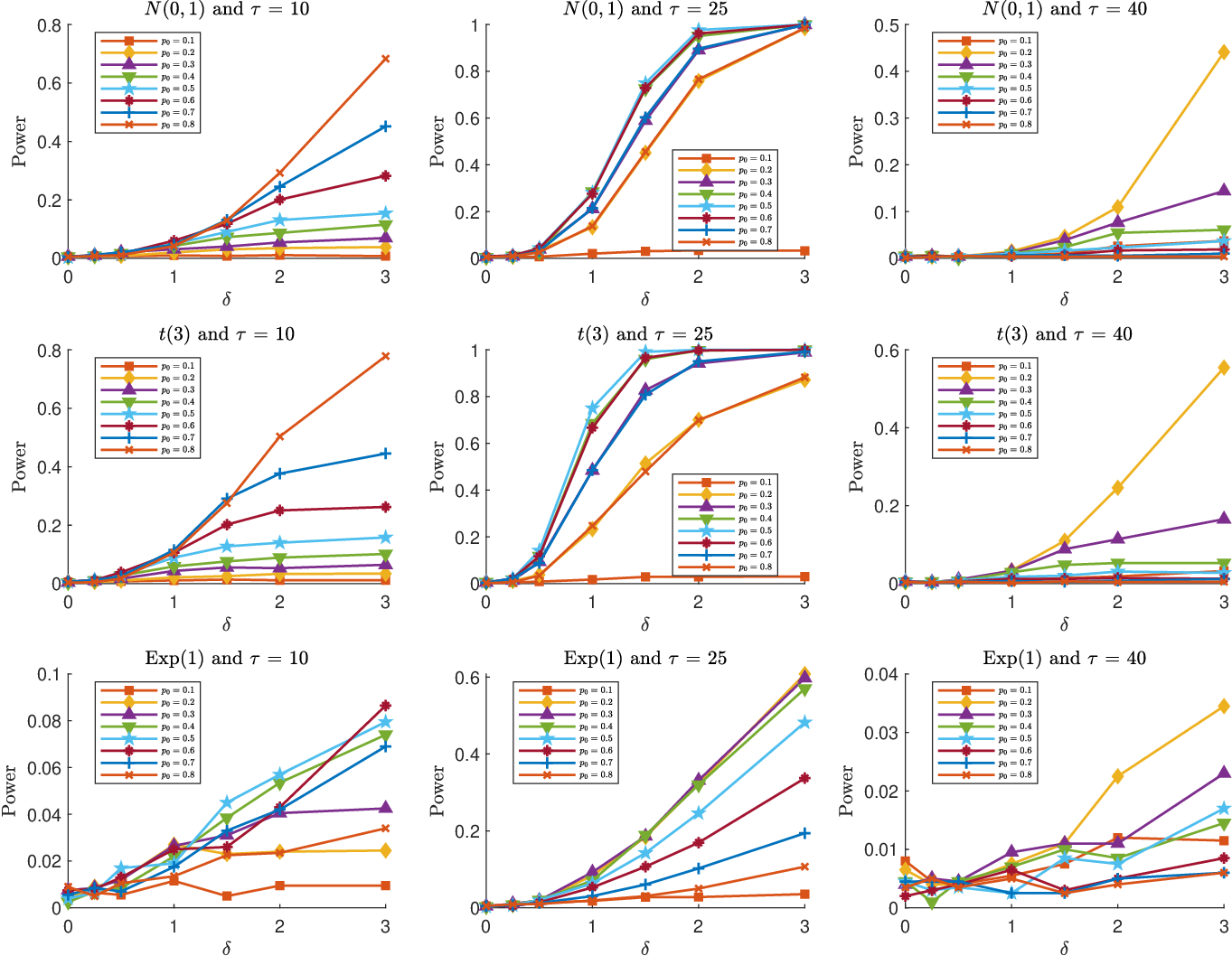}
 \caption{OC signal probabilities for the R-2 chart  with  $n=50$ and window size $r=25$.}\label{f2}
 \end{figure}

\begin{figure}[!h]
 \centering
 \includegraphics[scale=.44]{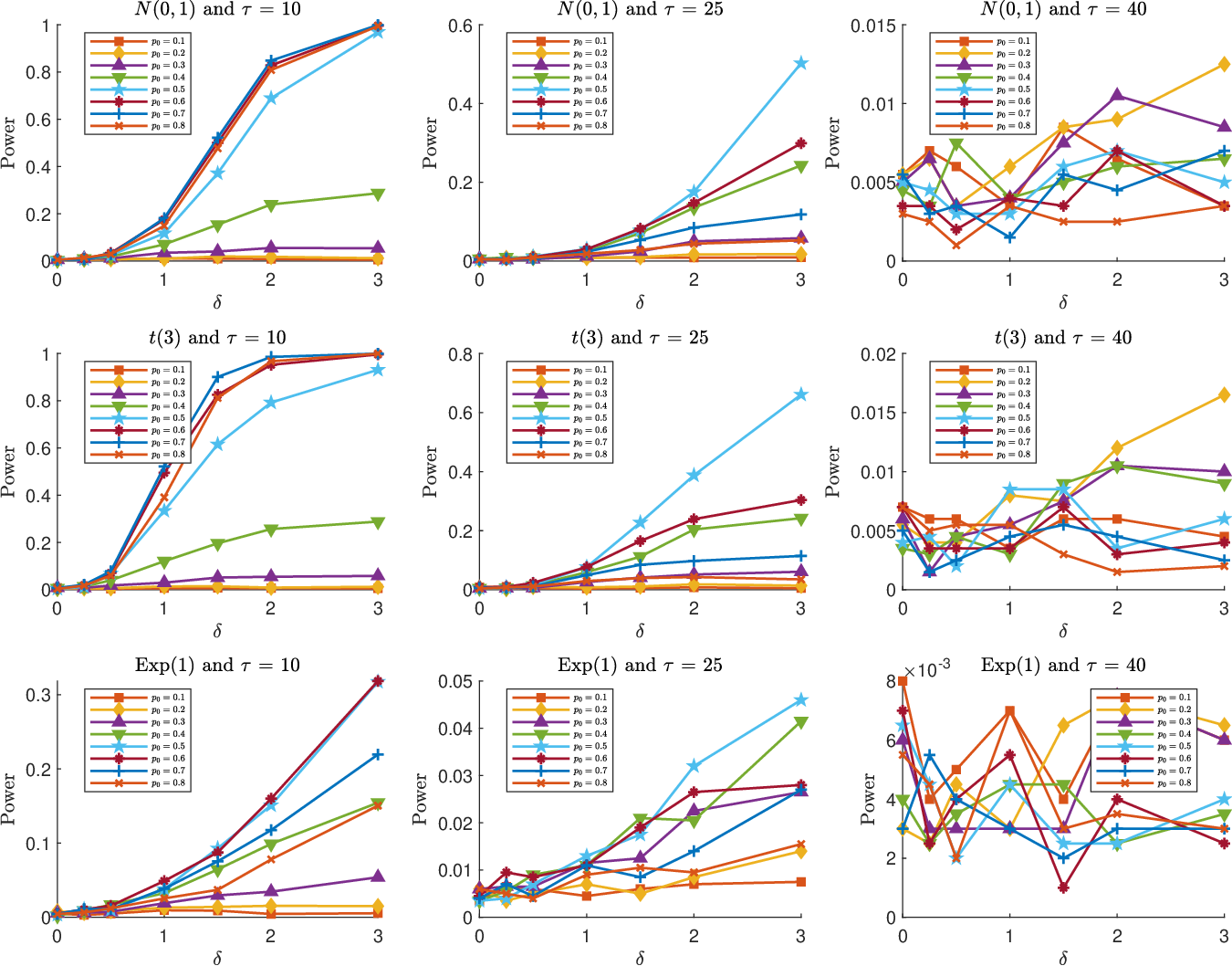}
 \caption{OC signal probabilities for the R-2 chart with $n=50$ and window size $r=40$.}\label{f3}
 \end{figure}

\begin{figure}[!h]
 	\centering
 	\includegraphics[scale=.44]{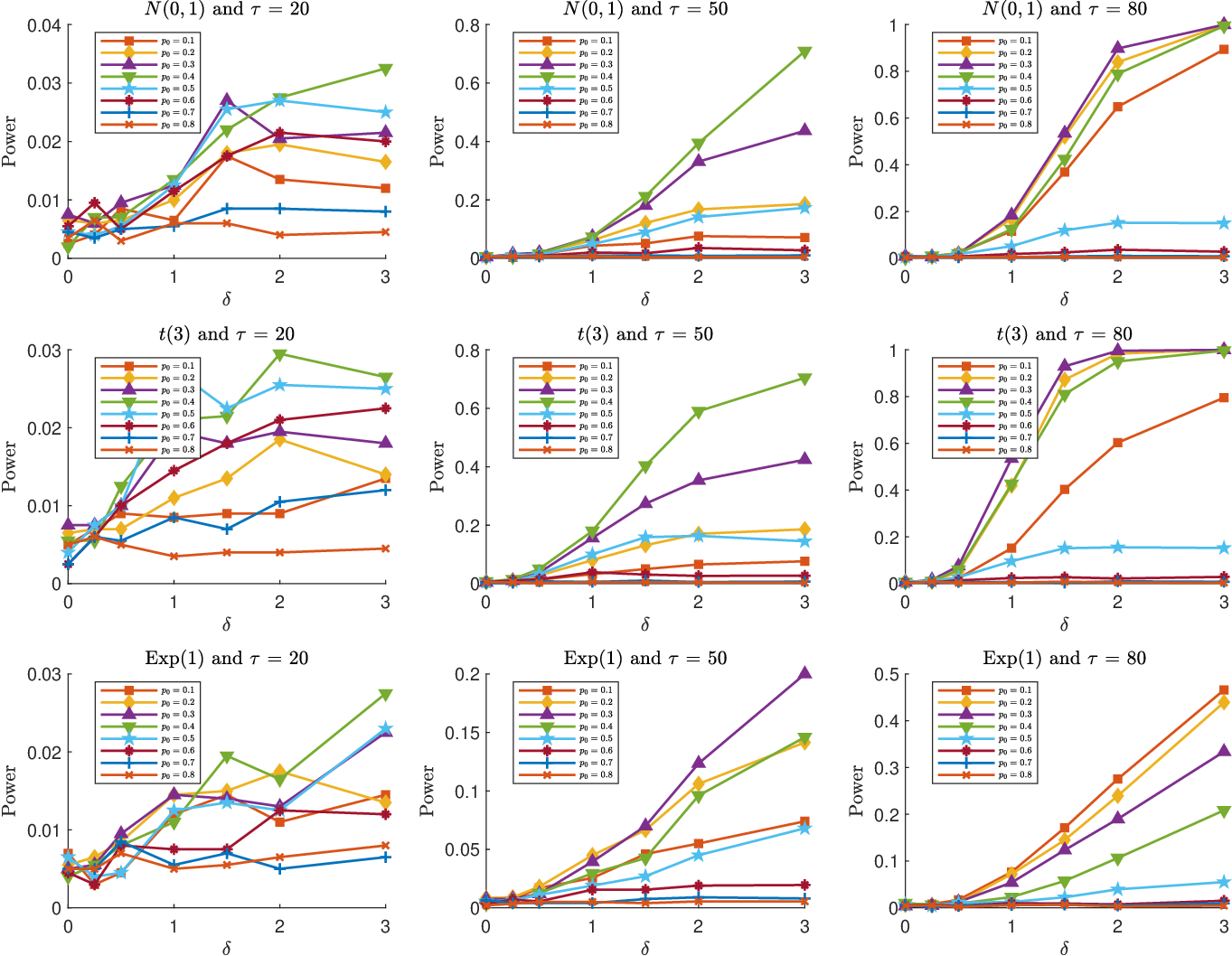}
 	\caption{OC signal probabilities for the R-2 chart with $n=100$ and window size $r=10$.}\label{f4}
 \end{figure}

\begin{figure}[!h]
 	\centering
 	\includegraphics[scale=.44]{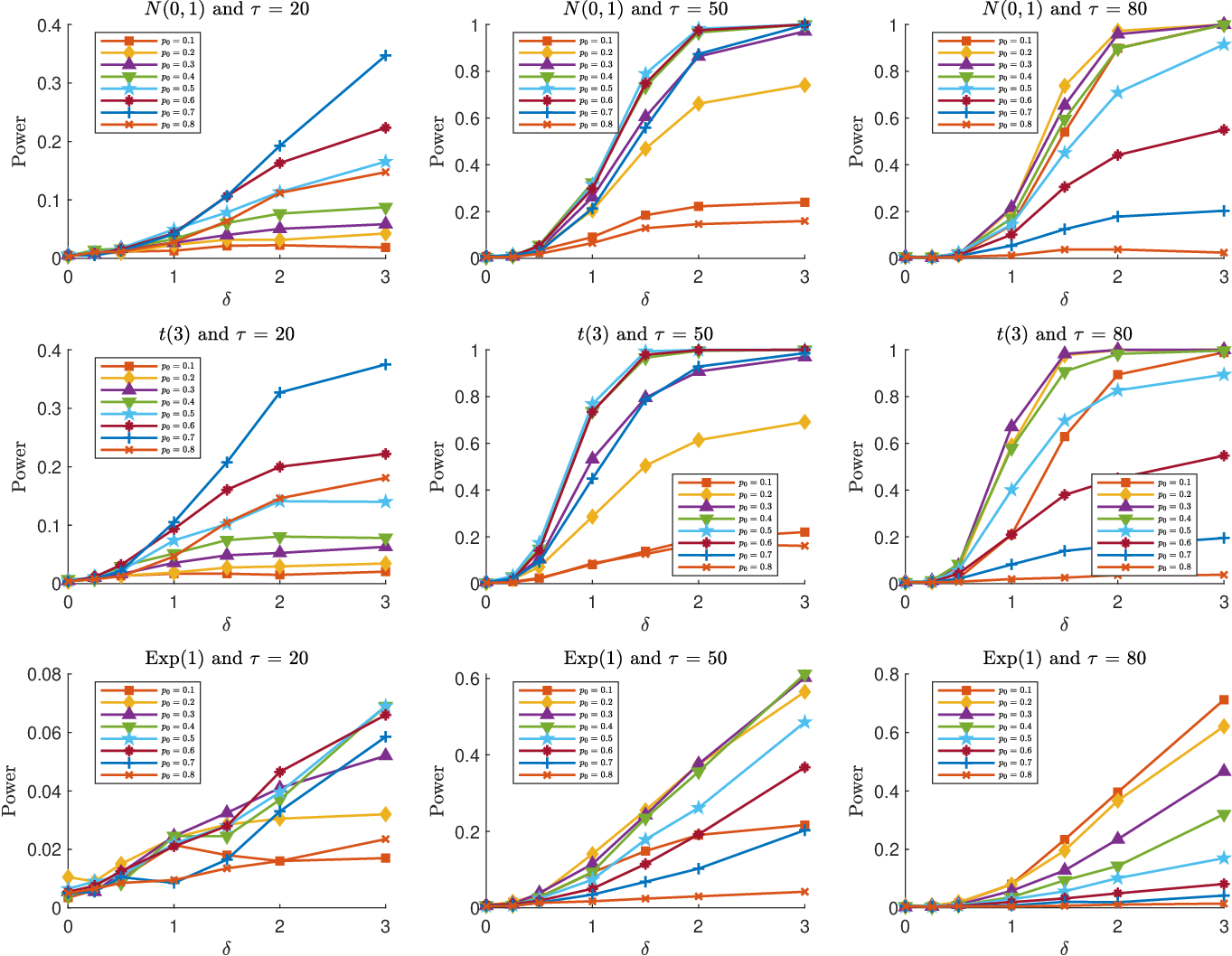}
 	\caption{OC signal probabilities for the R-2 chart with $n=100$ and window size $r=25$.}\label{f5}
 \end{figure}

\begin{figure}[!h]
 	\centering
 	\includegraphics[scale=.44]{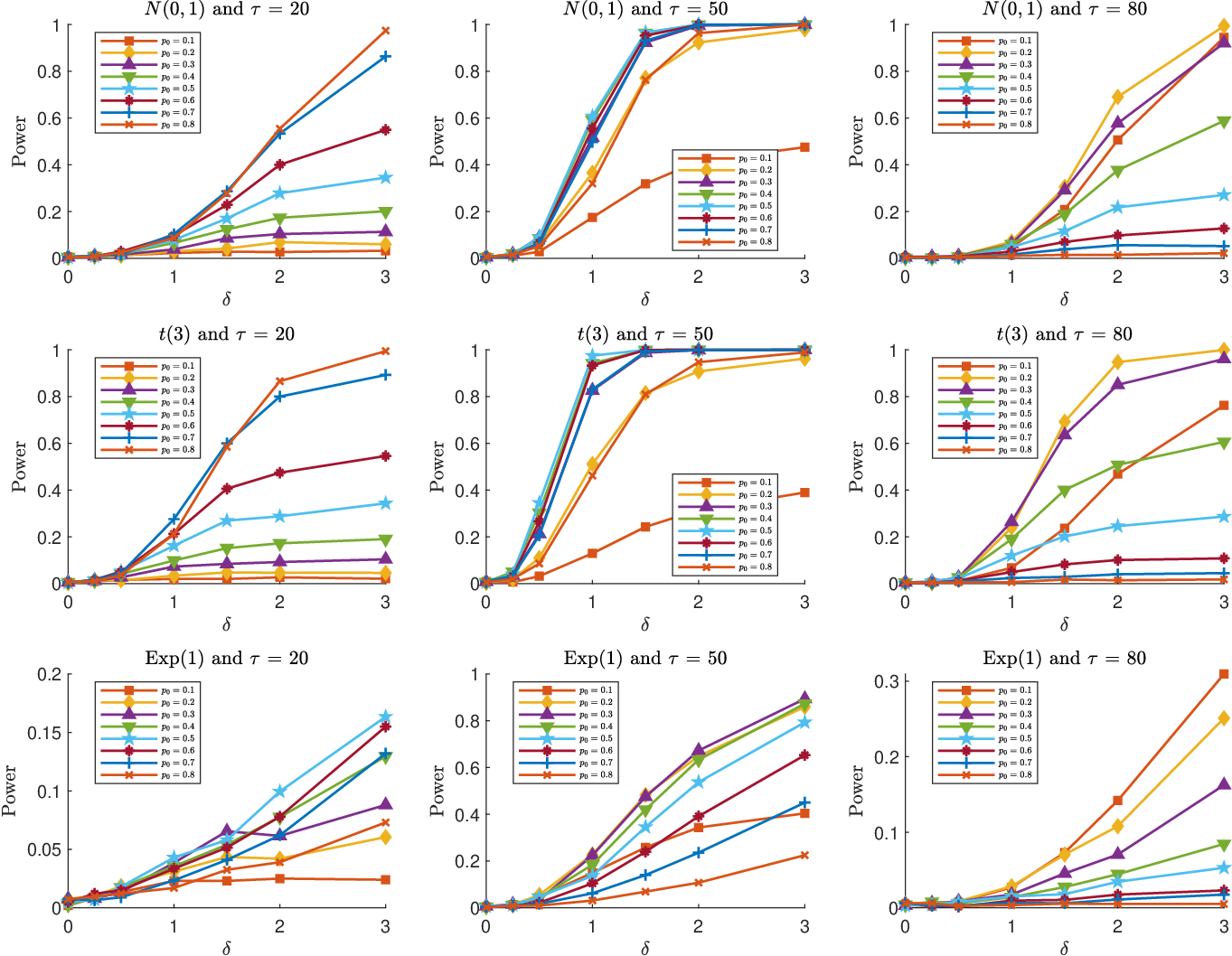}
 	\caption{OC signal probabilities for the R-2 chart with  $n=100$ and window size $r=40$.}\label{f6}
 \end{figure}

\begin{figure}[!h]
 \centering
 \includegraphics[scale=.44]{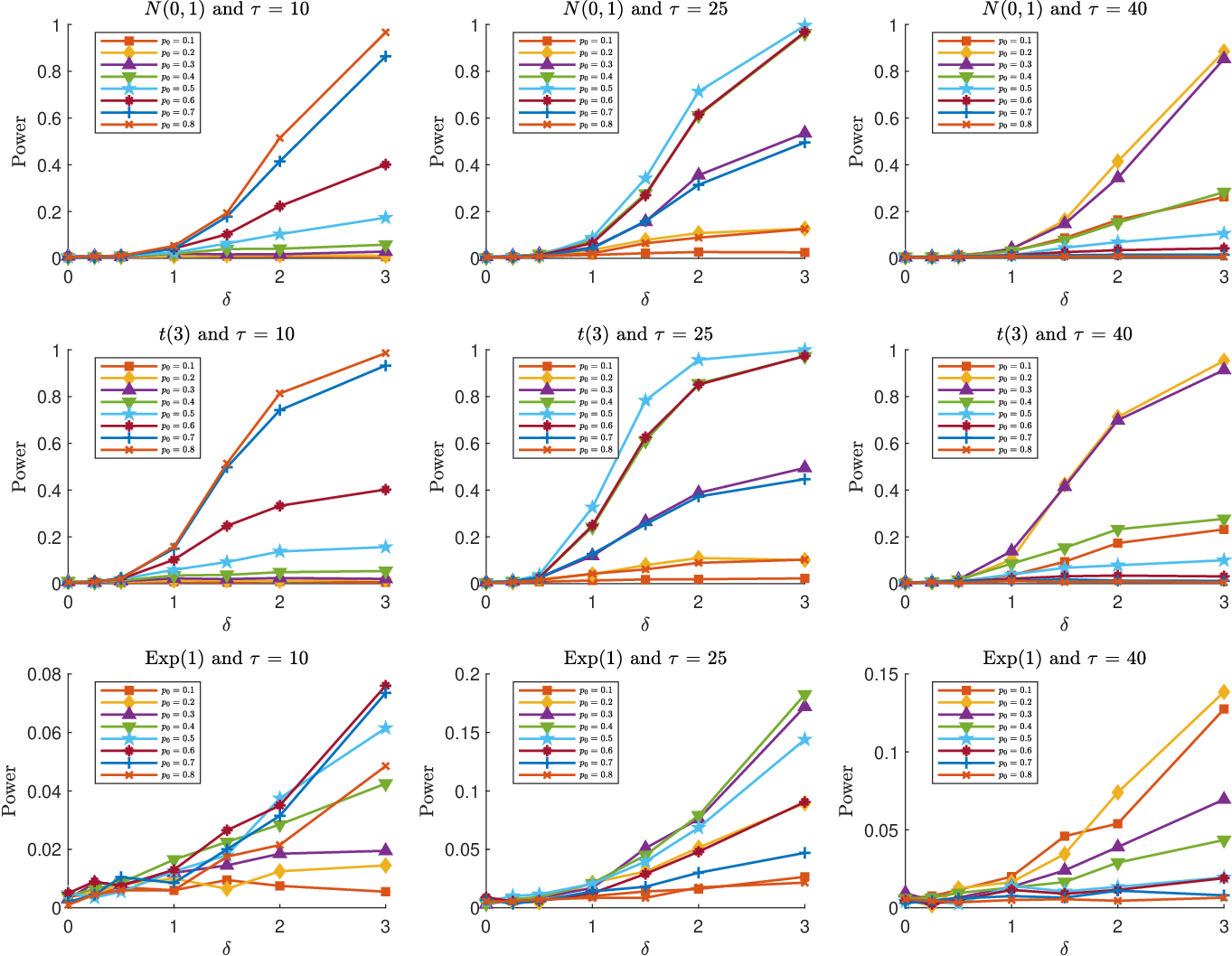}
 \caption{OC signal probabilities for the R-1 chart with $n=50$.}\label{f7}
 \end{figure}

\begin{figure}[!h]
 \centering
 \includegraphics[scale=.44]{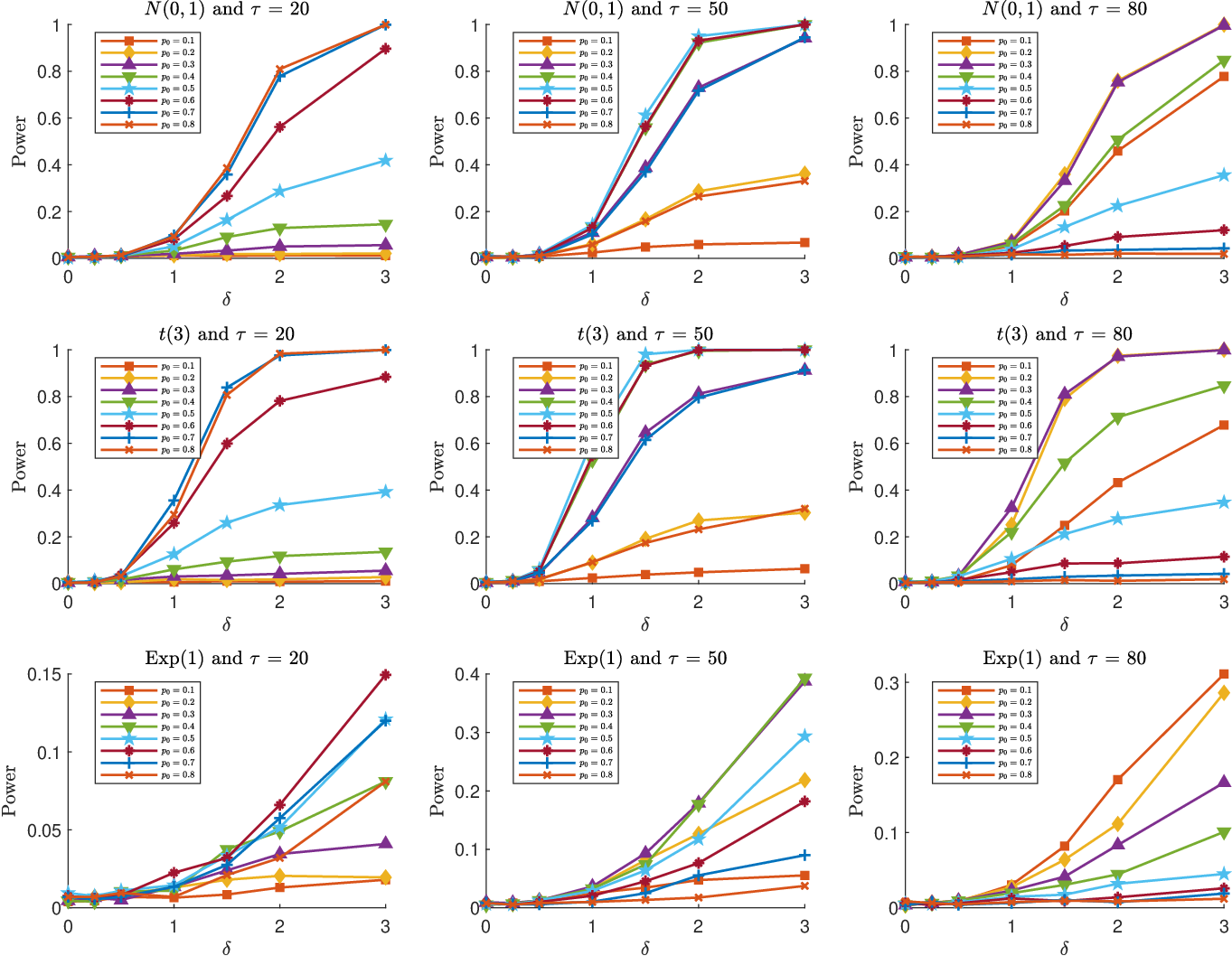}
 \caption{OC signal probabilities for the R-1 chart with  $n=100$.}\label{f8}
 \end{figure}

The empirical powers are evaluated under three shift scenarios and across three underlying distributions.
Figures~\ref{f1}--\ref{f6} display the performance of the proposed scan-based control charts, while Figures~\ref{f7} and~\ref{f8} illustrate the performance of the proposed run-based control charts.
Across all cases, the proposed methods maintain the nominal in-control signal probability and demonstrate stable out-of-control (OC) performance under diverse distributional conditions.

To preserve adequate resolution in the monitoring statistic, it is important to avoid selecting an excessively small or large reference proportion $p_0$.
Extreme choices of $p_0$ make the binary sequence overly sparse or overly dense, reducing the discriminative power of the chart.
The effect of the threshold $c$ (or equivalently, $p_0$) on chart performance can be observed in Figures~\ref{f1}--\ref{f6} for the scan-based charts.
In general, a smaller baseline proportion $p_0$ is preferable for shifts occurring later in the sequence, leading to fewer observed 1s and enhanced sensitivity near the end of the sequence.
Conversely, a larger $p_0$ is more effective for early location shifts, where a higher proportion of 1s promotes earlier detection of OC signals.
In addition, the choice of $p_0$ should account for the window size $r$.
As $r$ increases, the expected number of 1s per window rises, and a larger $p_0$ becomes advantageous to preserve contrast and detection power.
For smaller window sizes, a smaller $p_0$ should be adopted to prevent saturation in local counts.

The influence of distributional shape is also evident.
For symmetric distributions such as the normal, a balanced reference proportion ($p_0 \approx 0.5$) yields consistently high power and stable behavior across shift locations.
For right-skewed distributions such as the exponential, a smaller $p_0$ is recommended to offset the smaller probability of extreme high values.
By contrast, for heavy-tailed distributions such as the $t(3)$ distribution, a larger $p_0$ tends to enhance sensitivity, as more 1s are expected due to tail behavior.

We provide our recommendations for the choice of $p_0$ in Table~\ref{p0} across the proposed control charts and the three shift scenarios. The values in parentheses are those used for power comparisons with existing methods. Note that, as explained above, a smaller baseline proportion is required for right-skewed distributions; therefore, the baseline proportion is adjusted downward by 0.1 for $\text{Exp}(1)$.

      
 
      

 
  \begin{table}[!h]
 	%\caption{Global caption}2
 	\centering
 	\begin{tabular}{@{}|c|p{1.6cm}p{1.6cm}p{1.6cm}|@{}}
 		\hline
 		& Scenario I \,\,\,\,\, $\tau$ = 10 or 20 &Scenario II \,\,\, $\tau$ = 25 or 50& Scenario III \, $\tau$ = 40 or 80 \\\hline	
 		$r=10$  &0.4--0.6 (0.5) & 0.3--0.5 (0.4) &  0.1--0.3 (0.2)   	\\
 		$r=25$  &0.6--0.8 (0.7) & 0.4--0.6 (0.5) & 0.1--0.3 (0.2)	\\
 		$r=40$  &0.6--0.8 (0.7) & 0.4--0.6 (0.5) & 0.1--0.3 (0.2)	\\
 		$R_n$  &0.6--0.8 (0.7) & 0.4--0.6 (0.5) & 0.1--0.3 (0.2)	\\
 		\hline	
 	\end{tabular}
 	 	\caption{Recommended baseline proportions  $p_0$.}\label{p0}
 \end{table}

 To investigate the performance of our proposed control charts, we follow settings similar to those in \cite{Ning-2015}. Three existing nonparametric Phase-I control charts are examined for comparison with our proposed charts based on runs and patterns:
 \begin{description}
 	\item[1)] \textbf{Mann–Whitney (MW):} \cite{Hawkins-2010} proposed using the two-sample MW test for detecting location shifts without assuming normality. 	
 	
 	\item[2)] \textbf{Kolmogorov–Smirnov (KS):} Similar to the idea of \cite{Hawkins-2010}, Ross and Adams used the two-sample KS test to detect arbitrary distributional changes.
 	
 	\item[3)] \textbf{Empirical Likelihood Ratio (ELR):} \cite{Ning-2015} introduced a nonparametric Phase-I control chart for monitoring the location parameter, based on the maximum ELR statistic over potential change points. The null distribution of this statistic is approximated by a Gumbel distribution.
 \end{description}
 
\begin{figure}[!t]
 	\centering
 	\includegraphics[scale=.44]{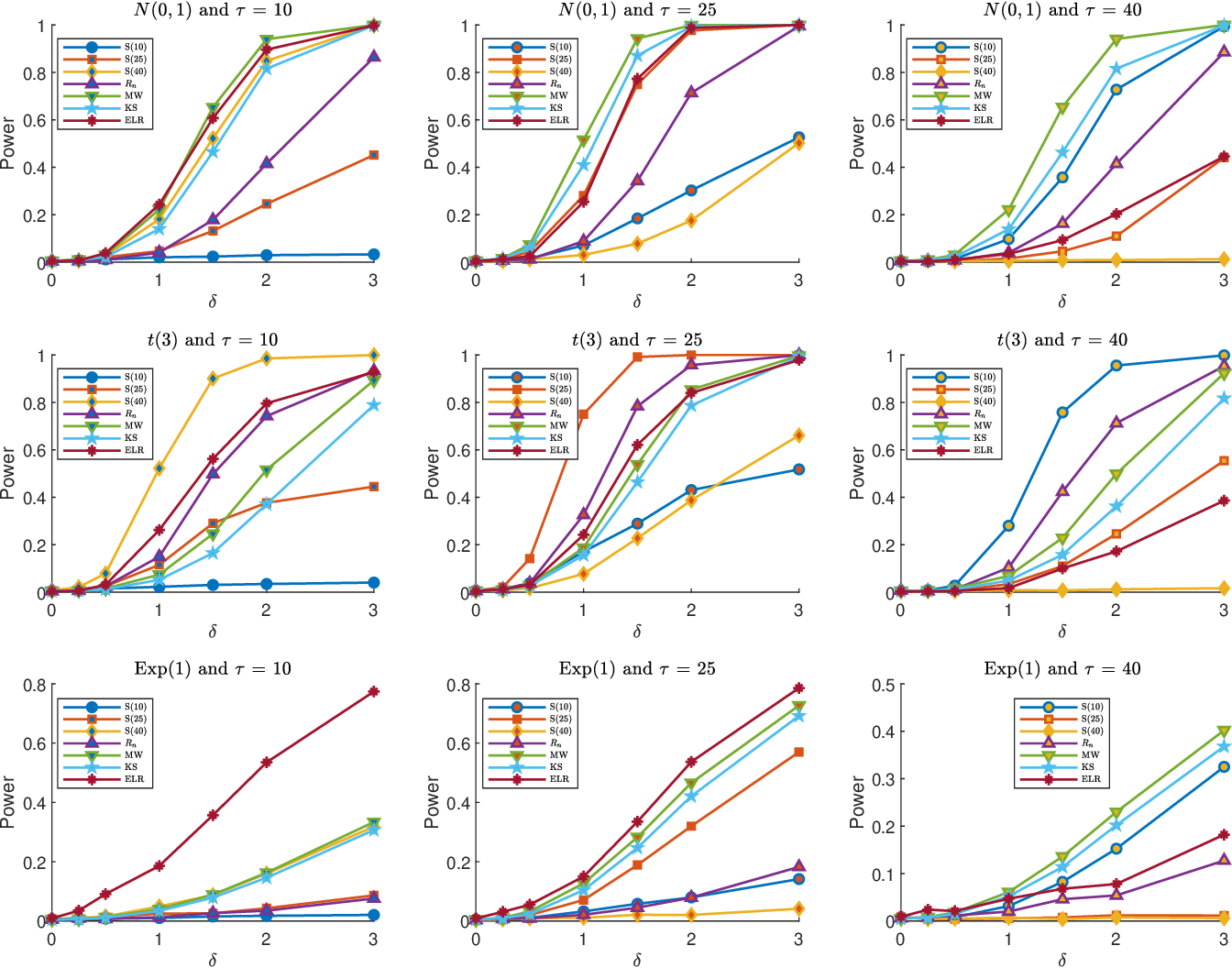}
 	\caption{OC signal probabilities for $n=50$. The signal probabilities of the MW, KS, and ELR charts are adopted from \cite{Ning-2015}.}\label{fig1}
 \end{figure}

\begin{figure}[!h]
 	\centering
 	\includegraphics[scale=.44]{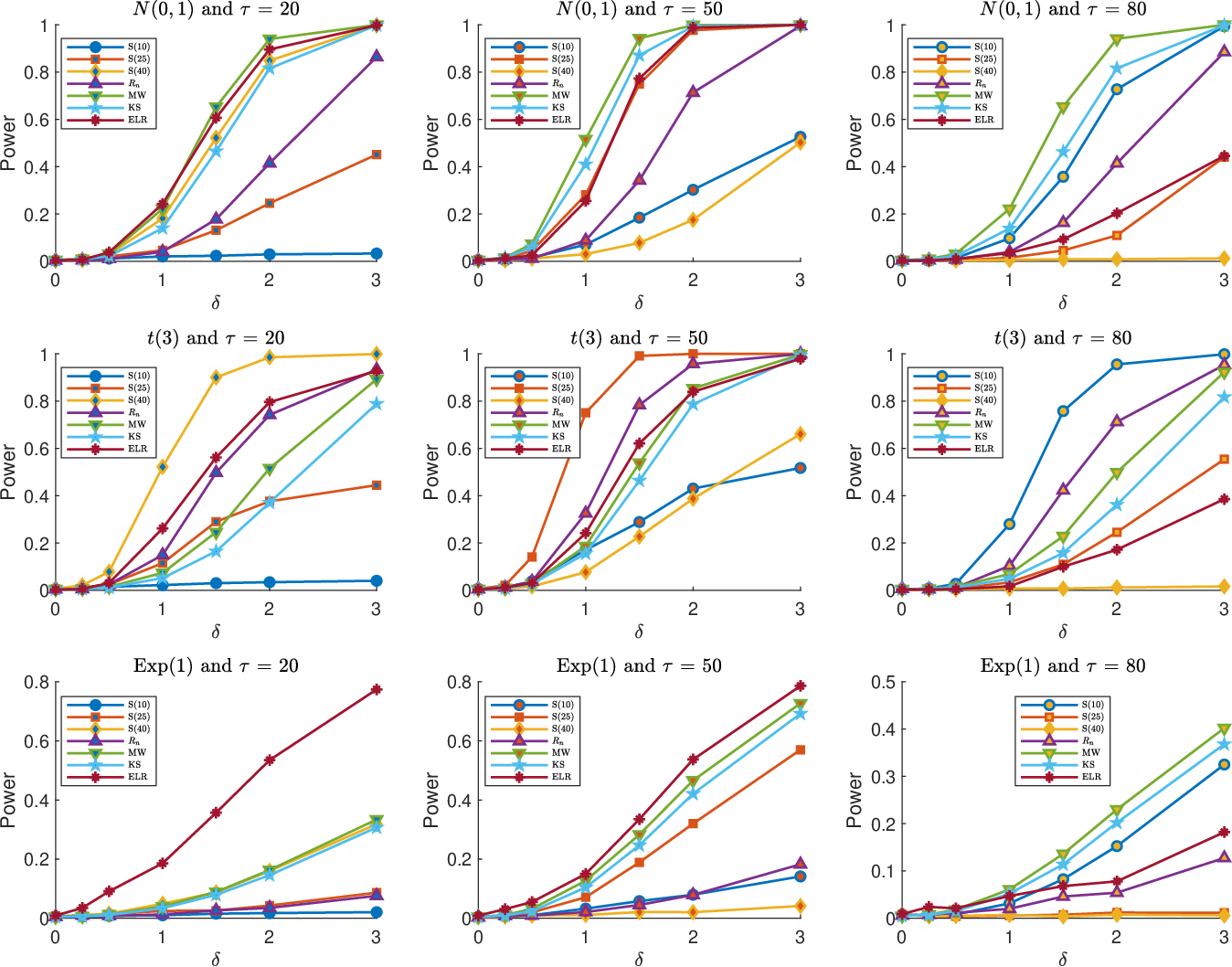}
 	\caption{OC signal probabilities for $n=100$. The signal probabilities of the MW, KS, and ELR charts are adopted from \cite{Ning-2015}.}\label{fig2}
 \end{figure}

 Table~\ref{t-0} presents the simulated IC signal probabilities, demonstrating that the proposed R-1 and R-2 control charts effectively control the IC signal probability at the desired level of 0.005. The OC signal probabilities are provided in Figures~\ref{fig1} and~\ref{fig2} for $n=50$ and $n=100$, respectively. From both Figures~\ref{fig1} and~\ref{fig2}, the other three control charts perform better than the proposed control charts for exponential distributions. However, our control charts perform satisfactorily for normal distributions and consistently better for heavy-tailed $t$-distributions. In particular, the OC signal probabilities of the R-1 chart based on the statistic $R_n$ tend to be higher when the change point occurs early or midway through the sample, regardless of whether the sample size is small or large. This characteristic is desirable, as it facilitates early detection of the change point. In contrast, the R-2 charts, which are based on scan statistics, generally yield higher OC signal probabilities when applied to heavy-tailed distributions such as the $t$-distribution. It is well known that scan statistics perform optimally when the scanning window size matches the actual cluster size. Therefore, it is unsurprising that the OC signal probabilities for $S_n(10)$, $S_n(25)$, and $S_n(40)$ are highest when the change points occur at $\tau=40$, $\tau=25$, and $\tau=10$, respectively. Among these, $S_n(25)$ and $S_n(40)$ generally outperform $S_n(10)$ when the sample size is $n=100$.

 \begin{table}[!t]
 	%\caption{Global caption}2
 	\centering
 	\begin{tabular}{@{}ccccccccc@{}}
 		\hline
 		& \multicolumn{2}{c}{$R_n$} & \multicolumn{2}{c}{$S_n(10)$}& \multicolumn{2}{c}{$S_n(25)$}& \multicolumn{2}{c}{$S_n(40)$}   \\
 		\cmidrule(l){2-3} \cmidrule(l){4-5}\cmidrule(l){6-7}\cmidrule(l){8-9}
 		&$n=50$ & $n=100$ &$n=50$ & $n=100$&$n=50$ & $n=100$&$n=50$ & $n=100$  \\\hline
 		$\mathcal{N}(0,1)$	& 0.0045 & 0.0055 &0.0051 & 0.0042 & 0.0043 & 0.0050 & 0.0053 & 0.0044		\\
 		Exp(1)  & 0.0057 & 0.0053 & 0.0042 & 0.0046 & 0.0039 & 0.0053 &   0.0054 & 0.0059\\
 		$t$(3)&0.0055 & 0.0056 & 0.0048 & 0.0045 & 0.0049 & 0.0058 & 0.0055 & 0.0048	 	\\
 		\hline	
 	\end{tabular}
 	 	\caption{IC signal probabilities.}\label{t-0}
 \end{table}
  
\section{An Application}

	We apply the proposed R-1 and R-2 control charts to the piston ring data presented in Tables 6.3 and 6E.7 of
\cite{Montgomery-2009}. 
The data in Table 6.3 contain 25 samples, each consisting of five inside diameter measurements of forged automobile engine piston rings. The $\bar{x}$ and $s$ control charts used in \cite{Montgomery-2009} indicate that the process is IC. There are an additional 15 samples for the piston ring process in Table 6E.7. Because subsamples were often collected at different production times, aggregating all 200 observations into a single Phase~I dataset may be problematic; therefore, we apply the charts to the 40 sample means to  determine whether a process instability is present.

We set up the control chart with an IC signal probability $\alpha=0.05$. The data exhibit  right skewness. The window sizes for the R-2 charts are selected to be 6 and 10. Thus, we choose small $p_0=0.2$ for $S_n(6)$, resulting in 8 ones and 32 zeros. A larger window size requires a somewhat larger $p_0$; therefore, we select $p_0=0.3$ for $S_n(10)$, resulting in 12 ones and 28 zeros.  The control limits (CLs) for the R-1 and R-2 charts are summarized in Table~\ref{t-2}. The desired signal probability may not be attained exactly since scan statistics are discrete.  The actual IC signal probabilities are given in parentheses.  

\begin{figure}[!htb]
	\centering
	\includegraphics[scale=.35]{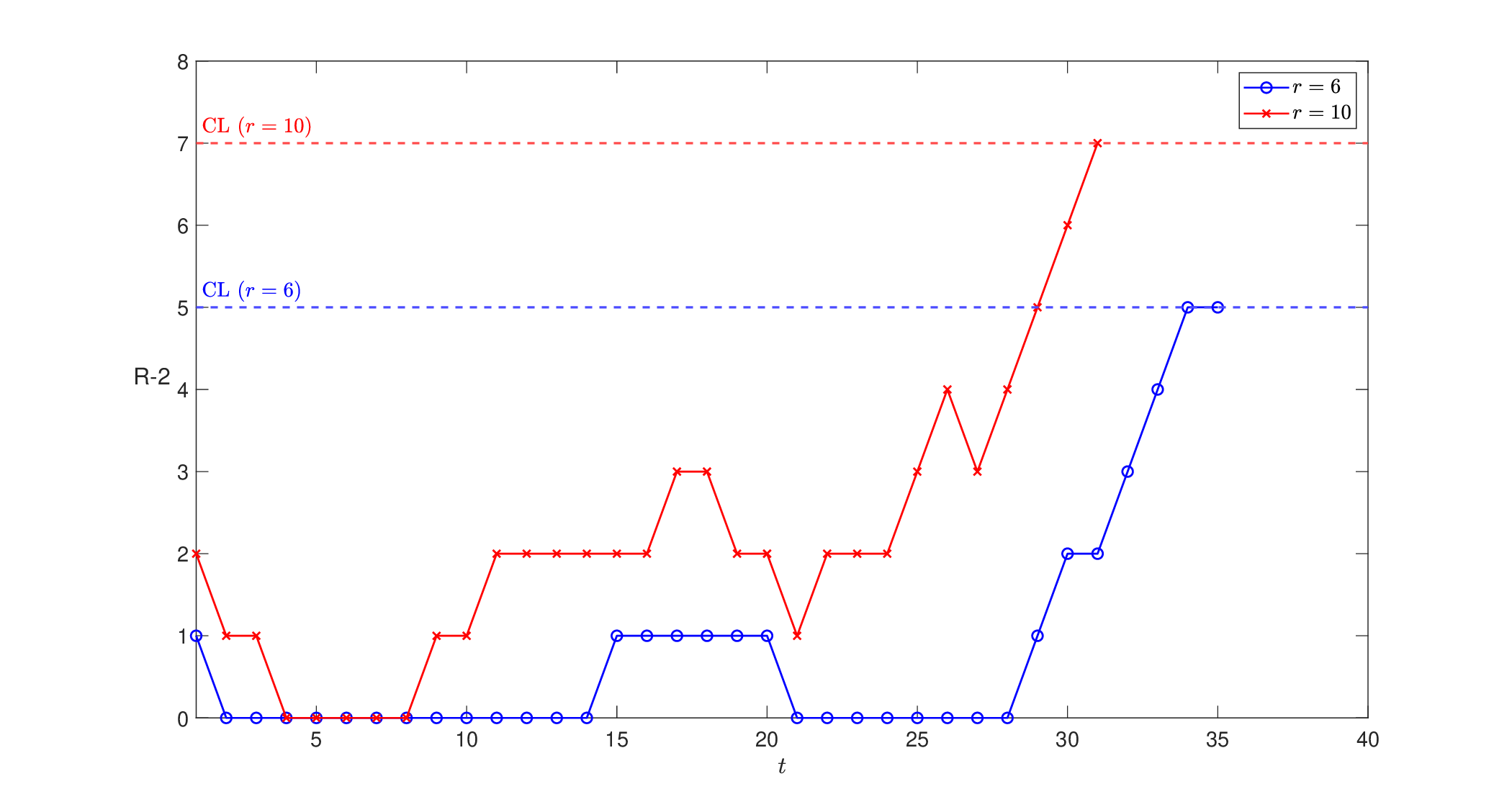}
	\caption{The R-2 chart of the piston ring data.}\label{f-3}
\end{figure}

The observed values are $R_{40} = 4$, $S_{40}(6) = 5$, and $S_{40}(10) = 7$. 
The corresponding window counts for the R-2 chart at each time point are shown in Figure~\ref{f-3}. 
Both the R-1 and R-2 charts signal a process mean shift at the 0.05 significance level. 
To further investigate potential sources of process instability, we list the time indices of windows with $p$-values smaller than 0.1. 
The scan statistics $S_{40}(6)$ and $S_{40}(10)$ identify $I_{34} \cup I_{35} = \{34, 35, 36, 37, 38, 39, 40\}$ ($p$-value = 0.0123) and $I_{31} = \{31, 32, 33, 34, 35, 36, 37, 38, 39, 40\}$ ($p$-value = 0.0525), respectively. 
Similarly, the run statistic $R_{40}$ indicates that the longest success run of 1s has length 4, occurring at $t = 37, 38, 39, 40$ ($p$-value = 0.0253). 
Thus, both charts consistently localize the same region as a potential source of process instability.

 \begin{table}[!t]
	%\caption{Global caption}2
	\centering
	\begin{tabular}{@{}cccc@{}}
		\hline
		& {$R_n(\alpha)$} & {$S_n(6,\alpha)$}& {$S_n(10,\alpha)$}   \\\hline	
		$\alpha=0.05$  &4 (0.0202, $p_0=0.2$) & 5 (0.0123, $p_0=0.2$)  & 7 (0.0525, $p_0=0.3$)  	\\
		\hline	
	\end{tabular}
		\caption{Control limits of the R-1 and R-2 control charts.}\label{t-2}
\end{table}

\section{Summary and Discussion}
 
This paper proposes a new class of distribution free control charts based on runs and patterns for individual observations. 
The methodology utilizes the conditional distributions of these statistics, particularly the number of success runs and the scan statistic, to design control charts that are independent of the underlying distribution.
The proposed approach leverages the finite Markov chain imbedding (FMCI) technique to obtain exact conditional distributions of the monitoring statistics given the number of successes in a Bernoulli sequence.

The strength of the proposed charts lies in their distribution-free nature, meaning that their in-control (IC) performance, such as the IC signal probability, does not depend on the underlying  distribution $F_0$. 
This property is especially advantageous in \emph{Phase~I} analysis, where the objective is to assess historical data to determine whether the process was IC without assuming a specific parametric form.

To implement the proposed method, the original observations are transformed into a Bernoulli sequence using an appropriate thresholding rule. 
This transformation can be based on a known or estimated reference distribution $F_0$, allowing the resulting sequence to be analyzed using runs- and patterns-type statistics. 
Among the statistics examined, the number of success runs and scan statistics were studied in detail, and their exact conditional distributions were derived using the FMCI technique. 
These conditional distributions were then employed to determine control limits for the proposed charts.

Several simulation studies were conducted to evaluate both in-control (IC) and out-of-control (OC) performance. 
The results demonstrate that the proposed charts maintain their nominal IC signal probability precisely and exhibit strong detection power across a range of shift scenarios for heavy-tailed distributions, where traditional parametric charts tend to be unreliable. 
Performance for skewed distributions was comparatively weaker for  the proposed and existing charts.

The selection of the baseline proportion $p_0$, which determines the threshold $c$ and thereby the balance of zeros and ones in the binary sequence, was also investigated. 
Simulation results indicate that moderate values of $p_0$ (typically between 0.2 and 0.7) provide stable performance across symmetric distributions and heavy-tailed distributions, while smaller $p_0$ values are preferable for right-skewed distributions such as the exponential. 
Recommended $p_0$ values for different window sizes and shift locations are summarized in Table~\ref{p0} to guide practical implementation.

Finally, the proposed R-1 and R-2 control charts were applied to the well-known piston ring data to demonstrate their practical utility. 
Both charts effectively detected a potential location shift. 
Specifically, the R-1 chart based on the number of success runs identified an extended sequence of large values near the end of the sample, while the R-2 chart based on scan statistics captured an early localized deviation in the sequence of sample means. 
Together, both charts provide a clearer diagnostic picture of process instability by pinpointing both the timing and extent of the abnormal region. These findings underscore the robustness and sensitivity of the proposed distribution-free control charts, demonstrating their capability to reveal subtle process shifts without relying on parametric assumptions.

In conclusion, the proposed framework provides a robust and flexible approach for \emph{Phase~I} process monitoring. 
The proposed control charts are truly distribution-free and flexible; other run or pattern statistics can also be utilized, allowing practitioners to select monitoring statistics suited to detecting specific types of shifts, such as variance changes, and to construct distribution-free control charts accordingly.
 
\bibliographystyle{spbasic}    % basic style, author-year citations
\bibliography{reference}   % name your BibTeX data base

\begin{thebibliography}{28}
\providecommand{\natexlab}[1]{#1}
\providecommand{\url}[1]{{#1}}
\providecommand{\urlprefix}{URL }
\expandafter\ifx\csname urlstyle\endcsname\relax
  \providecommand{\doi}[1]{DOI~\discretionary{}{}{}#1}\else
  \providecommand{\doi}{DOI~\discretionary{}{}{}\begingroup
  \urlstyle{rm}\Url}\fi
\providecommand{\eprint}[2][]{\url{#2}}

\bibitem[{Amin and Searcy(1991)}]{Amin-1991}
Amin RW, Searcy AJ (1991) A nonparametric exponentially weighted moving average
  control scheme. Communications in Statistics - Simulation and Computation
  20(4):1049--1072, \doi{10.1080/03610919108812996}

\bibitem[{Balakrishnan et~al(2010)Balakrishnan, Triantafyllou, and
  Koutras}]{Balakrishnan-2010}
Balakrishnan N, Triantafyllou IS, Koutras MV (2010) A distribution-free control
  chart based on order statistics. Communications in Statistics - Theory and
  Methods 39(20):3652--3677, \doi{10.1080/03610920903324858}

\bibitem[{Capizzi(2015)}]{Capizzi-2015}
Capizzi G (2015) Recent advances in process monitoring: Nonparametric and
  variable-selection methods for phase {I} and phase {II}. Quality Engineering
  27(1):44--67, \doi{10.1080/08982112.2015.968046}

\bibitem[{Chakraborti and Eryilmaz(2007)}]{Chakraborti-2007}
Chakraborti S, Eryilmaz S (2007) A nonparametric {S}hewhart-type signed-rank
  control chart based on runs. Communications in Statistics - Simulation and
  Computation 36(2):335--356, \doi{10.1080/03610910601158427}

\bibitem[{Chakraborti et~al(2001)Chakraborti, {Van Der Laan}, and
  Bakir}]{Chakraborti-2001}
Chakraborti S, {Van Der Laan} P, Bakir ST (2001) Nonparametric control charts:
  an overview and some results. Journal of Quality Technology 33(3):304--315,
  \doi{10.1080/00224065.2001.11980081}

\bibitem[{Chakraborti et~al(2009)Chakraborti, Eryilmaz, and
  Human}]{Chakraborti-2009}
Chakraborti S, Eryilmaz S, Human SW (2009) A phase {II} nonparametric control
  chart based on precedence statistics with runs-type signaling rules.
  Computational Statistics \& Data Analysis 53(4):1054--1065,
  \doi{10.1016/j.csda.2008.09.025}

\bibitem[{Champ and Woodall(1987)}]{Charles-1987}
Champ CW, Woodall WH (1987) Exact results for {S}hewhart control charts with
  supplementary runs rules. Technometrics 29(4):393--399,
  \doi{10.1080/00401706.1987.10488266}

\bibitem[{Chatterjee and Qiu(2009)}]{Chatterjee-2009}
Chatterjee S, Qiu P (2009) Distribution-free cumulative sum control charts
  using bootstrap-based control limits. The Annals of Applied Statistics
  3(1):349--369, \doi{10.1214/08-AOAS197}

\bibitem[{Chen et~al(2016)Chen, Zi, and Zou}]{Chen-2016}
Chen N, Zi X, Zou C (2016) A distribution-free multivariate control chart.
  Technometrics 58(4):448--459, \doi{10.1080/00401706.2015.1049750}

\bibitem[{Chowdhury et~al(2015)Chowdhury, Mukherjee, and
  Chakraborti}]{Chowdhury-2015}
Chowdhury S, Mukherjee A, Chakraborti S (2015) Distribution-free phase {II}
  {CUSUM} control chart for joint monitoring of location and scale. Quality and
  Reliability Engineering International 31(1):135--151, \doi{10.1002/qre.1677}

\bibitem[{Fu and Koutras(1994)}]{Fu-1994}
Fu JC, Koutras MV (1994) Distribution theory of runs: a {M}arkov chain
  approach. Journal of the American Statistical Association 89(427):1050--1058,
  \doi{10.1080/01621459.1994.10476841}

\bibitem[{Fu and Lou(2003)}]{Fu-lou-2003}
Fu JC, Lou WYW (2003) Distribution theory of runs and patterns and its
  applications. World Scientific, Singapore, \doi{10.1142/4669}

\bibitem[{Fu et~al(2002)Fu, Spiring, and Xie}]{Fu-2002}
Fu JC, Spiring FA, Xie H (2002) On the average run lengths of quality control
  schemes using a {M}arkov chain approach. Statistics \& Probability Letters
  56(4):369 -- 380, \doi{10.1016/S0167-7152(01)00183-3}

\bibitem[{Fu et~al(2003)Fu, Shmueli, and Chang}]{Fu-2003}
Fu JC, Shmueli G, Chang YM (2003) A unified chain approach for computing the
  run length distribution in control charts with simple or compound rules.
  Statistics \& Probability Letters 65(4):457 -- 466,
  \doi{10.1016/j.spl.2003.10.004}

\bibitem[{Fu et~al(2012)Fu, Wu, and Lou}]{Fu-Lou-Wu-2012}
Fu JC, Wu TL, Lou WYW (2012) Continuous, discrete, and conditional scan
  statistics. Journal of Applied Probability 49(1):199--209,
  \doi{10.1239/jap/1331216842}

\bibitem[{Hawkins and Deng(2010)}]{Hawkins-2010}
Hawkins DM, Deng Q (2010) A nonparametric change-point control chart. Journal
  of Quality Technology 42(2):165--173, \doi{10.1080/00224065.2010.11917814}

\bibitem[{Jones-Farmer et~al(2009)Jones-Farmer, Jordan, and
  Champ}]{Jones-Farmer-2009}
Jones-Farmer LA, Jordan V, Champ CW (2009) Distribution-free phase i control
  charts for subgroup location. Journal of Quality Technology 41(3):304--316,
  \doi{10.1080/00224065.2009.11917784}

\bibitem[{Jones-Farmer et~al(2014)Jones-Farmer, Ezell, and
  Hazen}]{Jones-Farmer-2014}
Jones-Farmer LA, Ezell JD, Hazen BT (2014) Applying control chart methods to
  enhance data quality. Technometrics 56(1):29--41,
  \urlprefix\url{http://www.jstor.org/stable/24587271}

\bibitem[{Koutras et~al(2007)Koutras, Bersimis, and Maravelakis}]{koutras-2007}
Koutras MV, Bersimis S, Maravelakis PE (2007) Statistical process control using
  {S}hewhart control charts with supplementary runs rules. Methodology and
  Computing in Applied Probability 9(2):207--224,
  \doi{10.1007/s11009-007-9016-8}

\bibitem[{Li et~al(2013)Li, Qiu, Chatterjee, and Wang}]{Li2013}
Li Z, Qiu P, Chatterjee S, Wang Z (2013) Using p values to design statistical
  process control charts. Statistical Papers 54(2):523--539,
  \doi{10.1007/s00362-012-0447-0}

\bibitem[{Lou(1997)}]{Lou-1997}
Lou WYW (1997) An application of the method of finite {M}arkov chain imbedding
  to runs tests. Statistics \& Probability Letters 31(3):155--161,
  \doi{10.1016/S0167-7152(96)00027-2}

\bibitem[{Montgomery(2009)}]{Montgomery-2009}
Montgomery D (2009) Introduction to statistical quality control, 6th edn.
  Wiley, Hoboken, NJ

\bibitem[{Ning et~al(2015)Ning, Yeh, Wu, and Wang}]{Ning-2015}
Ning W, Yeh AB, Wu X, Wang B (2015) A nonparametric phase {I} control chart for
  individual observations based on empirical likelihood ratio. Quality and
  Reliability Engineering International 31(1):37--55, \doi{10.1002/qre.1641}

\bibitem[{Parpoula(2021)}]{Parpoula-2021}
Parpoula C (2021) Phase {I} non-parametric control charts for individual
  observations: a selective review and some results. In: Dimotikalis Y,
  Karagrigoriou A, Parpoula C, Skiadas CH (eds) Applied Modeling Techniques and
  Data Analysis 1, Wiley, Hoboken, NJ, chap~13, pp 233--248,
  \doi{10.1002/9781119821588.ch13}

\bibitem[{Shmueli and Cohen(2003)}]{Shmueli-2003}
Shmueli G, Cohen A (2003) Run-length distribution for control charts with runs
  and scans rules. Communications in Statistics - Theory and Methods
  32(2):475--495, \doi{10.1081/STA-120018196}

\bibitem[{Wu(2020)}]{Wu-2020}
Wu TL (2020) Conditional waiting time distributions of runs and patterns and
  their applications. Annals of the Institute of Statistical Mathematics
  72(2):531--543, \doi{10.1007/s10463-018-0696-3}

\bibitem[{Zhou et~al(2009)Zhou, Zou, Zhang, and Wang}]{Zhou-2009}
Zhou C, Zou C, Zhang Y, Wang Z (2009) Nonparametric control chart based on
  change-point model. Statistical Papers 50(1):13--28,
  \doi{10.1007/s00362-007-0054-7}

\bibitem[{Zou and Tsung(2010)}]{Zou-2010}
Zou C, Tsung F (2010) Likelihood ratio-based distribution-free {EWMA} control
  charts. Journal of Quality Technology 42(2):174--196,
  \doi{10.1080/00224065.2010.11917815}

\end{thebibliography}

\end{document}